\documentclass{appolb}
\usepackage{graphicx}

\usepackage[T1]{fontenc}
\usepackage[utf8]{inputenc}
\usepackage[lowtilde]{url}

\usepackage{amsmath}
\usepackage{mathtools}
\usepackage{microtype}

\usepackage{floatflt}

\begin{document}
\title{Constraining PDFs and nPDFs with recent data%
\thanks{Presented at the 29th International Conference on Ultra-relativistic Nucleus-Nucleus Collisions, April 4-10, 2022, Krak\'ow, Poland.}%
}
\author{Petja Paakkinen
\address{\centering University of Jyväskylä, Department of Physics\\ P.O. Box 35, FI-40014 University of Jyväskylä, Finland\\ and\\
Helsinki Institute of Physics\\ P.O. Box 64, FI-00014 University of Helsinki, Finland}
}
\date{July 28, 2022}
\maketitle
\begin{abstract}
The progress in constraining proton and nuclear parton distribution functions is briefly summarised. Some persistent uncertainties are pointed out and  recent experimental advancements highlighted.
\end{abstract}

\section{Introduction}

In many applications of high-energy hadronic collisions, including those of heavy nuclei, the partonic structure of hadrons manifests itself through the collinear factorisation of QCD~\cite{Collins:1989gx}. That is, the cross section for producing an inclusive final state $k+X$, where the $X$ denotes ``anything'', can be described in terms of coefficient functions (CF) ${\rm d}\hat{\sigma}^{ij \rightarrow k+X'}$, which are calculable from perturbative QCD, and parton distribution functions (PDFs) $f_i^A, f_j^B$, which contain long-range physics and thus cannot be obtained by perturbative means, plus ``higher twist'' (HT) corrections which are suppressed at high enough momentum-transfer scale $Q \gg \Lambda_{\rm QCD}$. Formally,
\begin{equation}
  {\rm d}\sigma^{AB \rightarrow k+X} (Q^2) \!\overset{Q \gg \Lambda_{\rm QCD}}{=}\! \sum_{i,j,X'} f_i^A(Q^2) \otimes {\rm d}\hat{\sigma}^{ij \rightarrow k+X'}(Q^2) \otimes f_j^B(Q^2) + {\cal O}(1/Q^2),
\end{equation}
where ``$\otimes$'' indicates a multiplicative (Mellin) convolution
over $x$, the fraction of the parents momentum carried by the parton.

The PDFs $f_i^A(x,Q^2)$, with $i$ denoting the parton flavour and $A$ the parent hadron or nucleus, exhibit two important properties: First, they are universal, independent of the partonic process (i.e.\ the CFs). And second, their scale dependence is governed by the DGLAP equations
$\partial f_i^A / \partial \ln Q^2 = \sum_j P_{ij} \otimes f_j^A$ with perturbatively calculable splitting functions $P_{ij}$. For these properties, it is possible to extract the PDFs from global analyses of experimental data of multiple observables at different scales.

\begin{figure}[tb]
\centerline{%
 \includegraphics[width=\textwidth]{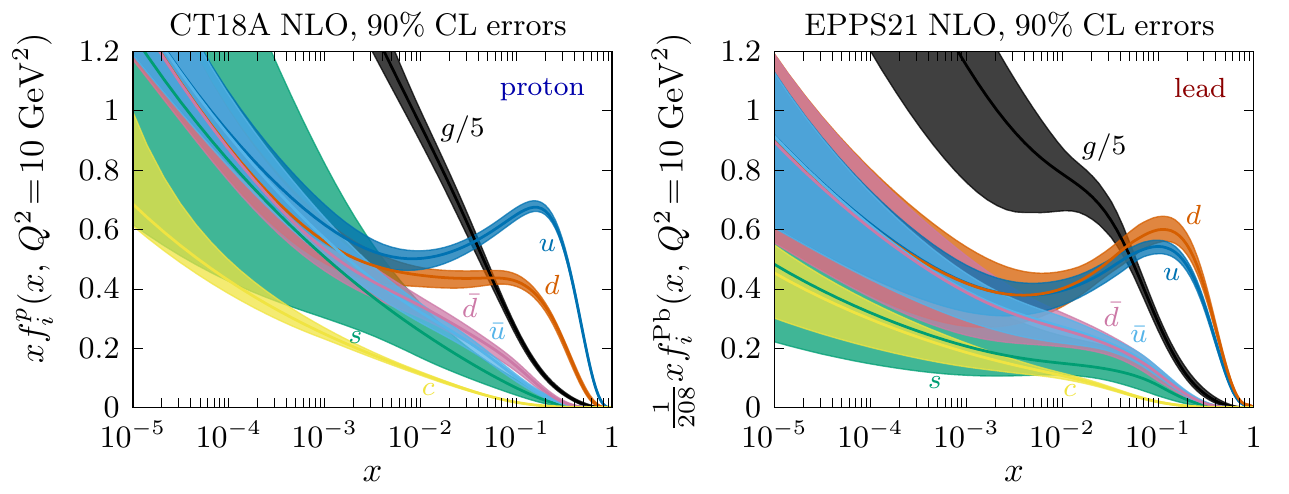}
}
\vspace{-0.1cm}
\caption{A representative example of proton and nuclear PDFs taken from the CT18A NLO proton-PDFs~\cite{Hou:2019efy} and EPPS21 NLO nPDFs~\cite{Eskola:2021nhw}. The coloured bands show the estimated PDF uncertainties at 90\% confidence level.}
\label{Fig:abs_PDFs}
\end{figure}

Fig.~\ref{Fig:abs_PDFs} shows results from such analyses with the CT18A next-to-leading order (NLO) proton PDFs~\cite{Hou:2019efy} and EPPS21 NLO nuclear PDFs (nPDFs)~\cite{Eskola:2021nhw}, where the latter uses the former as its free-proton baseline, taken as representative examples. As is common, both analyses fit six independent parton flavours, $u, d, \bar{u}, \bar{d}, s=\bar{s}$ and glue, taking heavy-quark PDFs fully perturbatively generated.\footnote{Some analyses, like the recent NNPDF4.0 extraction~\cite{NNPDF:2021njg}, extend this set to fit also intrinsic charm and strange asymmetry.} The general features are evident from this plot: For protons, most flavours are already quite well constrained in a wide range of $x$. The exception to this rule is the strange quark, which is currently the least constrained flavour in both free-proton and nuclear PDF analyses. For nPDFs, the uncertainties persist to be larger compared to the free-proton ones, particularly in the small-$x$ region where direct data constraints are scarce.

\vspace{-0.5cm}
\section{Progress in proton-PDF analyses}

The most recent proton-PDF analyses use several thousand data points from various experiments. Such a global approach is a necessity: a multi-observable fit is needed in order to constrain individual flavours and a multi-experiment fit in order to do so in a wide range of $x$. For example, fixed target deep-inelastic scattering (DIS) and Drell-Yan (DY) data are important in setting the large-$x$ quark distributions (valence-to-sea and flavour separation). While much of these data are already rather old, new data are still coming from Fermilab and JLab, where e.g.\ the new SeaQuest Collaboration measurement of proton-deuteron to proton-proton ratio of DY dilepton production~\cite{SeaQuest:2021zxb} improves our understanding of the antimatter asymmetry in the proton. To reach smaller values of $x$, collider DIS data are needed, and nowadays the HERA inclusive and heavy-quark data~\cite{H1:2015ubc,H1:2018flt} form the backbone of any modern proton-PDF analysis. With the large $x, Q^2$ lever arm, these data are able to constrain also the gluon distribution through the DGLAP equations. Recently, however, most of the new data have come from hadron colliders (in particular from the CERN LHC), granting access to a wide spectrum of new processes such as the production of electroweak (EW) bosons, jets, top-pairs, etc., allowing for testing the factorisation and PDF flavour dependence in new kinematic windows.

One of the most important latest developments in this field has been the inclusion of next-to-next-to-leading-order (NNLO) corrections~\cite{Currie:2016bfm,Currie:2017eqf} to the perturbative jet-production calculations in the most recent global analyses.\footnote{At leading colour, that is. Calculations at full colour have been performed only very recently~\cite{Czakon:2019tmo,Chen:2022tpk}.} These jet observables, together with the $t\bar{t}$ production, are important in setting the large-$x$ gluon distributions~\cite{NNPDF:2021njg}. The NNLO precision improves fit quality particularly for the LHC (EW boson and jet) data, and for example the recent MSHT20 analysis~\cite{Bailey:2020ooq} reports on an improvement of 700 units in $\chi^2_{\rm tot}$ for the 4363 global data points. Therefore, with the present LHC data precision, NNLO proton-PDFs are no longer a mere formal improvement, reducing theoretical uncertainties from missing-higher-order corrections, but a necessity for an accurate fit.

\begin{floatingfigure}[r]{0.55\textwidth}
\vspace{-0.35cm}
\centerline{%
 \includegraphics[width=0.55\textwidth]{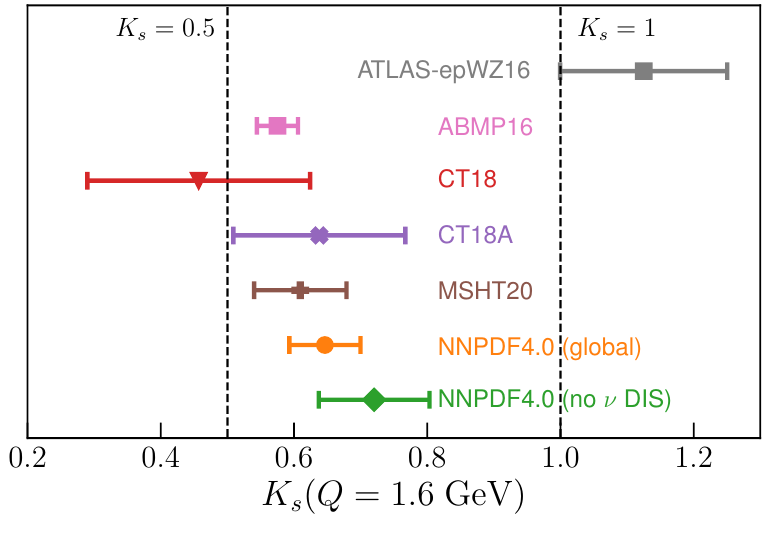}
}
\vspace{-0.1cm}
\caption{Strangeness suppression in different proton-PDF analyses. Figure from Ref.~\cite{Feng:2022inv}.}
\label{Fig:proton_strangeness}
\end{floatingfigure}

LHC data can also challenge some of the earlier approaches taken in global analyses. Traditionally, proton-PDF fits have included
neu\-tri\-no-nu\-cle\-us ($\nu A$) DIS data for improved strange-quark constraints, which have led to suppressed strangeness e.g.\ in the ABMP16 fit~\cite{Alekhin:2017kpj}, as is quantified by the fraction
\begin{equation}
  K_s = \frac{\int_0^1 {\rm d}x x[{s}(x) \!+\! {\bar{s}}(x)]}{\int_0^1 {\rm d}x x[{\bar{u}}(x) \!+\! {\bar{d}}(x)]}
\end{equation}
shown in Fig.~\ref{Fig:proton_strangeness}. Complementary data from ATLAS EW-boson production have then confronted this view with a preference for unsuppressed strange (see ATLAS-epWZ16 in Fig.~\ref{Fig:proton_strangeness})~\cite{ATLAS:2012sjl,ATLAS:2016nqi}. Finding some tension between these two data types, the CT18~\cite{Hou:2019efy} analysis provides two different fits, where the nominal CT18 set does not include the ATLAS data, but the CT18A version does. Later, a simultaneous fit was found to be feasible when NNLO charm-quark mass corrections in the $\nu A$ cross sections were accounted for, as in the nNNPDF4.0 and MSHT20 analyses~\cite{NNPDF:2021njg,Bailey:2020ooq}.

All of this points to the direction that understanding the role of nuclear corrections in fitting proton PDFs with nuclear data is becoming increasingly important. Indeed, the recent NNPDF4.0 analysis finds very different large-$x$ sea-quark behaviour depending on whether the uncertainties from nNNPDF2.0 nPDFs were included or not~\cite{NNPDF:2021njg}, in contrast to some earlier evaluations~\cite{Ball:2018twp}. How to take these nuclear corrections into account varies from analysis to analysis, the MSHT20 taking them from the DSSZ nPDFs~\cite{deFlorian:2011fp} plus an additional three-parameter fit~\cite{Bailey:2020ooq}, whereas CT18 uses a scale-independent phenomenological
parametrisation as described in Ref.~\cite{Accardi:2021ysh}.

\vspace{-0.3cm}
\section{Progress in nPDF analyses}

The nPDFs are fitted with similar global analyses as their free-proton counterparts, relying only to the QCD collinear factorisation and thus giving a model-agnostic way to study the nuclear effects. Unlike in proton-PDF analyses, where NNLO is the standard, most of the nPDF analyses are still performed at NLO and those recent analyses which also provide NNLO fits~\cite{Khanpour:2020zyu,Helenius:2021tof} are performed with a rather limited number of data types and thus cannot parametrise the flavour dependence very freely. Those (NLO) analyses standing on a more global footing instead include a variety of data from the LHC proton-lead (p+Pb) runs, with EPPS21~\cite{Eskola:2021nhw} and nNNPDF3.0~\cite{AbdulKhalek:2022fyi} also propagating the uncertainties from their free-proton baselines.

A very important advancement in this field has been the inclusion of the double-differential LHC p+Pb dijet and D$^0$-production data in the recent analyses, leading to strong new constraints on the nuclear gluon PDFs~\cite{Eskola:2021nhw,AbdulKhalek:2022fyi} which have been previously poorly known in lack of collider DIS data on nuclei.\footnote{The heavy-quark data were recently considered also in Ref.~\cite{Duwentaster:2022kpv}.} In accordance with the earlier reweighting study~\cite{Eskola:2019dui}, these analyses obtain a good fit to the dijet data, except in the very forward bins where the predictions overshoot the data. This could be due to the missing data correlations, but also larger than expected NNLO or non-perturbative corrections could play a role. For the D$^0$-production, nNNPDF3.0, with its POWHEG+PYTHIA approach, finds a large scale uncertainty in the nuclear ratio and therefore fit only to the more constraining forward data, whereas EPPS21, with the S-ACOT-$m_{\rm T}$ mass scheme~\cite{Helenius:2018uul}, finds no large theory uncertainty in this ratio in the fitted $p_{\rm T} > 3\ {\rm GeV}$ region~\cite{Eskola:2019bgf}. Complementary gluon constraints can be derived from single-inclusive production of $\pi^0$, $\pi^\pm$ and $K^\pm$ as in the nCTEQ15WZSIH analysis~\cite{Duwentaster:2021ioo} fitting to a large set of data from PHENIX, STAR and ALICE measurements. In such fits, like with the D$^0$-production, one is sensitive to the fragmentation functions and their associated uncertainties, which however partially cancel in the nuclear ratios.

In addition, $W^\pm$ production and the most precise CMS Collaboration measurement from Run~2~\cite{CMS:2019leu} in particular, have been already included in almost every recent nPDF fit. The approaches taken in doing so, however, vary, TUJU21~\cite{Helenius:2021tof}, nNNPDF3.0~\cite{AbdulKhalek:2022fyi} and nCTEQ15WZSIH~\cite{Duwentaster:2021ioo} fitting to absolute cross sections, whereas EPPS21~\cite{Eskola:2021nhw} using nuclear-modification ratios to cancel proton-PDF uncertainties~\cite{Eskola:2022rlm}.
Also new $Z$-boson data from Run~2 recently became available~\cite{CMS:2021ynu} and have been studied in several nPDF analyses, all finding some difficulties in describing the data. While the discrepancy in the small-invariant-mass bin can be explained by NNLO corrections~\cite{Helenius:2021tof,AbdulKhalek:2022fyi}, the on-peak deviations appear to be due to rather abrupt change in the shape of the data at midrapidity, possibly due to untypically large data fluctuations~\cite{Eskola:2021nhw}.

\begin{floatingfigure}[r]{0.53\textwidth}
\vspace{-0.5cm}
\centerline{%
 \includegraphics[width=0.53\textwidth]{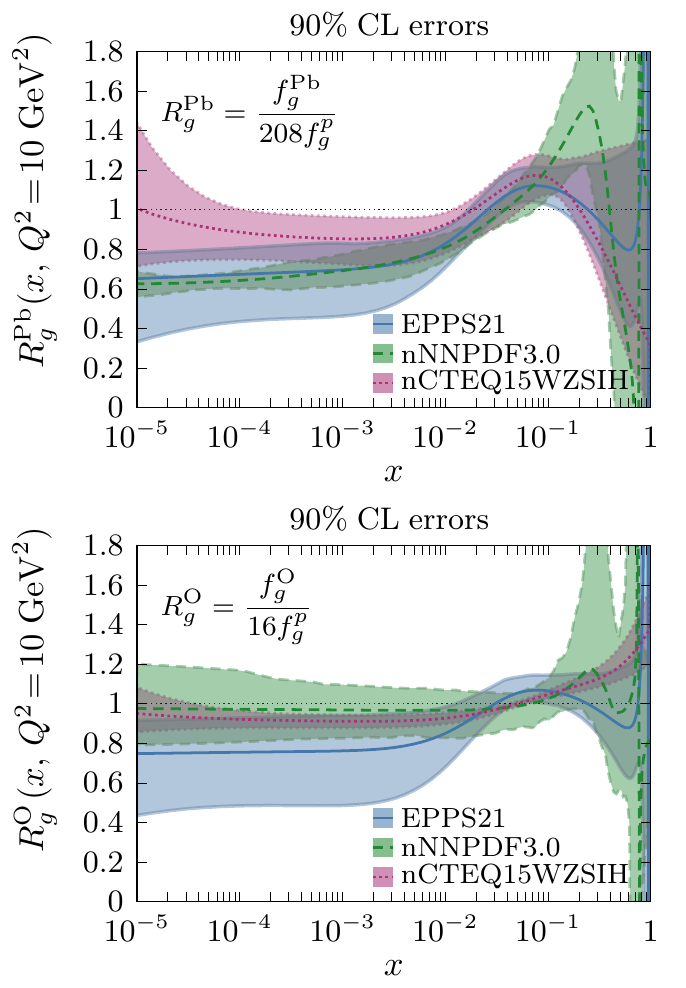}
}
\caption{Nuclear modifications of the gluon PDFs in lead (top) and oxygen (bottom) nuclei. Results from Refs.~\cite{Eskola:2021nhw,AbdulKhalek:2022fyi,Duwentaster:2021ioo}.}
\label{Fig:comp_nPDFs}
\end{floatingfigure}

The resulting gluon distributions from the EPPS\-21, nNNPDF\-3.0 and nCTEQ\-15\-WZ\-SIH analyses~\cite{Eskola:2021nhw,AbdulKhalek:2022fyi,Duwentaster:2021ioo} are compared in Fig.~\ref{Fig:comp_nPDFs}. All of these nPDFs exhibit small-$x$ gluon shadowing and mid-$x$ antishadowing in lead, which are now established from multiple experimental observables. In finer details, the results however differ. At small $x$, EPPS\-21 and nNNPDF\-3.0 agree extremely well due to the use of D$^0$ and dijet data, though with rather different uncertainty estimations, whereas nCTEQ\-15\-WZ\-SIH with the main small-$x$ constraints coming from EW-boson production favouring smaller shadowing. At large $x$, nNNPDF\-3.0 shows a different behaviour from EPPS\-21 which now agrees better with nCTEQ\-15\-WZ\-SIH instead. This is possibly associated with the PHENIX $\pi^0$ data included in EPPS\-21 and nCTEQ\-15\-WZ\-SIH and also with the exclusion of backward D$^0$ data from nNNPDF\-3.0. It should be also noted that the discussed direct gluon constraints come only from data on heavy nuclei and therefore the gluon distributions in lighter nuclei, such as oxygen, remain poorly constrained~\cite{Paakkinen:2021jjp}.

The progress does of course not end here. In Ref.~\cite{Shen:2021eir} the advantages of measuring triple-differential dijet production in p+Pb were discussed. Since the triple-differential measurement fixes partonic kinematics at leading order, this would give a powerful test of factorisation and PDFs. Also the new $\pi^0$-production small-system scan from PHENIX~\cite{PHENIX:2021dod} with improved precision and access to higher $p_{\rm T}$ can help in deciphering the nuclear modifications at large $x$. Contrary to nPDF expectations, the measured ``Cronin peak'' size follows an ordering $^3{\rm He}$+${\rm Au} < d$+${\rm Au} < p$+${\rm Au}$ and Ref.~\cite{PHENIX:2021dod} discusses the cause of these effects in terms of multiple-scattering (i.e.\ HT) and flow-like effects. In this conference, also new LHCb measurements of D$^0$s~\cite{LHCb:2022rlh} and $\pi^0$s~\cite{LHCb:2022vfn} at 8.16 TeV and charged hadrons at 5.02 TeV~\cite{LHCb:2021vww} in p+Pb were presented, probing smaller values of $x$. The hopes for using exclusive $J/\psi$ photoproduction in constraining nPDFs were also elevated with the first phenomenological implementation of the NLO corrections in ultrapheripheral Pb+Pb~\cite{Eskola:2022vpi}. A more direct probe of the collinearly factorised nPDFs in ultrapheripheral events is the inclusive dijet photoproduction, for which the ATLAS measurement at 5.02 TeV has now been finally fully unfolded~\cite{ATLAS:2022cbd}.

\vspace{-0.15cm}
\section{Conclusion}

The new data constraints from LHC have brought the proton PDFs (and increasingly also the nPDFs) to the point where taking NNLO corrections into account is a necessity for a good fit, and the p+Pb data have also put unprecedented constraints on the gluon nPDFs. There is also ongoing work to understand the (cross)correlations between proton and nuclear PDF analyses, and many new data sets (with the LHC Run~3 also underway) to be exploited in the future.

\bigskip
I thank T.~J.~Hobbs and P.~M.~Nadolsky for clarifying the use of nuclear corrections in CT18. Financial support from the Academy of Finland, project 330448, is acknowledged. This research was funded as a part of the Center of Excellence in Quark Matter of the Academy of Finland and as a part of the European Research Council project ERC-2018-ADG-835105 YoctoLHC. Figs.~\ref{Fig:abs_PDFs} and~\ref{Fig:comp_nPDFs} use protanopia and deuteranopia safe colour schemes from Refs.~\cite{Wong:2011} and~\cite{Tol:2021}.

\vspace{-0.1cm}
\bibliographystyle{IEEEtran}
\bibliography{APPB_QM2022_Paakkinen.bib}

\end{document}